\documentclass[iop, twocolappendix, tighten]{emulateapj}
\usepackage{apjfonts}
\usepackage{verbatim}
\usepackage{xcolor}
\usepackage{natbib}
\usepackage{amssymb,amsmath,mathrsfs}
\tabletypesize{\scriptsize}

\usepackage{color,hyperref}
\definecolor{linkcolor}{rgb}{0,0,0.5}
\hypersetup{colorlinks=true,linkcolor=linkcolor,citecolor=linkcolor,
            filecolor=linkcolor,urlcolor=linkcolor}
\usepackage{url}
\usepackage{amssymb,amsmath}
\usepackage{subfigure}

\def\h2{$\rm H_2$}

\newcommand{\msun}{M$_{\odot}$}

\newcommand{\halpha}{H$\alpha$}

\newcommand{\hi}{H{\sc I}}

\newcommand{\lsun}{L$_{\odot}$}
\newcommand{\tauninety}{$\tau_{90}$}
\newcommand{\logavgmass}{log($\langle$M$_{\star}$/M$_{\odot}$$\rangle$)}
\newcommand{\rvir}{R$_{\rm virial}$}

\begin{document}

\title{The Star Formation Histories of Local Group Dwarf Galaxies \sc{III}. \\ Characterizing Quenching in Low-Mass Galaxies\altaffilmark{*}}

\author{
Daniel R.\ Weisz\altaffilmark{1,6},
Andrew E.\ Dolphin\altaffilmark{2}, 
Evan D.\ Skillman\altaffilmark{3},
Jon Holtzman$^{4}$,
Karoline M. Gilbert\altaffilmark{5},
Julianne J.\ Dalcanton\altaffilmark{1},
Benjamin F. Williams\altaffilmark{1}
}

\altaffiltext{*}{Based on observations made with the NASA/ESA Hubble Space Telescope, obtained from the Data Archive at the Space Telescope Science Institute, which is operated by the Association of Universities for Research in Astronomy, Inc., under NASA constract NAS 5-26555.}
\altaffiltext{1}{Department of Astronomy, University of Washington, Box 351580, Seattle, WA 98195, USA; dweisz@uw.edu}
\altaffiltext{2}{Raytheon Company, 1151 East Hermans Road, Tucson, AZ 85756, USA}
\altaffiltext{3}{Minnesota Institute for Astrophysics, University of Minnesota, 116 Church Street SE, Minneapolis, MN 55455, USA}
\altaffiltext{4}{Department of Astronomy, New Mexico State University, Box 30001, 1320 Frenger St., Las Cruces, NM 88003}
\altaffiltext{5}{Space Telescope Science Institute, 3700 San Martin Drive, Baltimore, MD, 21218, USA}
\altaffiltext{6}{Hubble Fellow}

\begin{abstract}

We explore the quenching of low-mass galaxies (10$^4$ $\lesssim$ M$_{\star}$ $\lesssim$ 10$^8$ M$_{\odot}$) as a function of lookback time using the star formation histories (SFHs) of 38 Local Group dwarf galaxies. The SFHs were derived from analyzing color-magnitude diagrams of resolved stellar populations in archival Hubble Space Telescope/Wide Field Planetary Camera 2 imaging.  We find: (1) Lower mass galaxies quench earlier than higher mass galaxies; (2) Inside of \rvir\ there is no correlation between a satellite's current proximity to a massive host and its quenching epoch; (3) There are hints of systematic differences in quenching times of M31 and Milky Way (MW) satellites, although the sample sample size and uncertainties in the SFHs of M31 dwarfs prohibit definitive conclusions. Combined with literature results, we qualitatively consider the redshift evolution ($z=0-1$) of the quenched galaxy fraction over $\sim$ 7 dex in stellar mass (10$^4$ $\lesssim$ M$_{\star}$ $\lesssim$ 10$^{11.5}$ \msun).  The quenched fraction of all galaxies generally increases toward the present, with both the lowest and highest mass systems exhibiting the largest quenched fractions at all redshifts.  In contrast, galaxies between M$_{\star}$ $\sim$ 10$^8$--10$^{10}$ \msun\ have the lowest quenched fractions.  We suggest that such intermediate-mass galaxies are the least efficient at quenching.  Finally, we compare our quenching times with predictions for infall times of low-mass galaxies associated with the MW.  We find that some of the lowest-mass satellites (e.g., CVn {\sc II}, Leo {\sc IV}) may have been quenched before infall while higher mass satellites (e.g., Leo {\sc I}, Fornax) typically quench $\sim$ 1-4 Gyr after infall.

\end{abstract}


\section{Introduction}
\label{sec:intro}

Environment is believed to play an important role in the evolution of low-mass galaxies.  Their shallow potentials make them vulnerable to a variety of internal (e.g., galactic winds, supernovae feedback) and external (e.g., tidal effects, ram pressure stripping) processes that can reshape and remove their baryons.  The central role of environment is observationally borne out by the dwarf `morphology-density' relationship, which shows that red, gas-poor, non-star forming dwarf galaxies are predominantly found in close proximity to a massive host, while blue, gas-rich, star-forming dwarfs are preferentially located in lower density environments \citep[e.g.,][]{hodge1971, einasto1974, vandenbergh1994, mateo1998, vandenbergh2000, grcevich2009, tolstoy2009, weisz2011a, geha2012, mcconnachie2012, spekkens2014}.  This powerful observational constraint favors models in which quenching (i.e., the complete cessation of star formation)\footnote{We define quenched galaxies to be those without cold gas and without evidence of recent star formation (e.g., as traced by \halpha\ and ultra-violet flux).  In these systems, the re-ignition of star formation is unlikely without significant gas accretion.  We view such quenched systems as qualitatively different from post-starburst galaxies, despite the use of similar `quenching' terminology in the literature.  Generally, post-starburst galaxies have had recent but no current star formation (i.e., ultra-violet flux but no detectable \halpha), and typically have large gas reservoirs, suggesting that they may be capable of restarting star formation in the future.} is induced by environmental mechanisms \citep[e.g.,][]{mayer2001, mayer2006} as opposed to purely internal processes \citep[e.g.,][]{dekel1986}.

Despite the clear importance of environment for gas removal and quenching in low-mass galaxies, our knowledge of the responsible mechanisms is less certain.  Tidal effects and ram pressure are believed to be the primary mechanisms for shutting down star formation \citep[e.g.,][]{gunn1972, einasto1974, faber1983}. However, their relative effectiveness is modulated by a number of factors including halo mass and accretion history, infall time, proximity to a host, stellar feedback (which can induce outflows and weaken the effective potential), and the heating of gas from cosmic reionization in the lowest mass systems \citep[e.g.,][]{dekel1986, thoul1996, gnedin2000, mayer2001, busha2010, governato2010, zolotov2012, lu2014a, wetzel2015}.  Because the complex interplay between each of these processes can vary with time, the morphology-density relationship at $z=0$ cannot strongly discriminate between various scenarios for creating environmentally dependent quenching in dwarf galaxies.

Orthogonal constraints on quenching mechanisms come from star formation histories (SFHs).  By tracing star formation over cosmic time, it is possible to determine exactly when and how quickly a dwarf galaxy quenched, providing insights into the physical mechanisms responsible for shutting down star formation.  SFHs for galaxies located within a few Mpc can be reconstructed in detail by analyzing their resolved star color-magnitude diagrams \citep[CMDs; e.g.,][]{tosi1989, harris2001, dolphin2002, aparicio2009}.  The power of SFHs for constraining quenching was recently illustrated by \citet{sohn2013} in their analysis of  Milky Way (MW) satellite Leo {\sc I}.  The reconstructed orbital history has revealed that Leo {\sc I} entered the virial radius of the MW 2 Gyr ago and made a pericentric passage $\sim$ 1 Gyr ago at a distance of 90 kpc.  The SFH shows an enhancement followed by complete cessation of star formation over the same timescale, suggestive of a rapid quenching scenario likely due to ram pressure stripping, as its pericentric distance seems too large for tidal effects to be significant.

Despite the clear utility of SFHs for quenching studies, there have been few efforts to systematically characterize and analyze quenching in nearby dwarfs.  The majority of relevant studies largely focus on quantifying the effects of early reionization on the lowest mass (M$_{\star}$ $\lesssim$ 10$^{6}$ \msun) MW satellites \citep[e.g.,][]{sand2009, sand2010, okamoto2012, brown2012, brown2014}.  Outside of the MW sub-group, there is only a small sample of quenched, low-mass galaxies with well-constrained SFHs \citep[Cetus, Tucana, Andromeda {\sc II}, Andromeda {\sc XVI};][]{monelli2010b, monelli2010c, weisz2014m31}, making it challenging to explore broad quenching trends in low-mass systems.  At the same time several studies have compared N-body simulations to $z=0$ properties of low-mass galaxies to constrain quenching timescales \citep[e.g.,][]{slater2013, slater2014, phillips2014, wheeler2014, phillips2015}, but combining SFHs with the simulations is still an emerging field \citep[e.g.,][]{rocha2012}.

In this paper, we use the SFHs of 38 LG dwarf galaxies to systematically measure the quenching epochs and timescales in low-mass systems (M$_{\star}$ $\lesssim$ 10$^8$ \msun).  The SFHs are taken from the \citet{weisz2014a} uniform analysis of optical CMDs based on archival imaging taken with the Hubble Space Telescope / Wide Field Planetary Camera 2 \citep[HST/WFPC2;][]{holtzman1995}.  Using these SFHs, we investigate empirical trends in quenching epochs and timescales as functions of basic galaxy properties (e.g., stellar mass, proximity to a host), place the LG into the context of broader studies of quenching, and compare our data with select models of satellite infall in the LG.  We have previously used this dataset to examine the observational evidence for reionization as a quenching mechanism in the lowest mass systems, and refer readers to \citet{weisz2014b} for a more detailed discussion of this specific quenching mechanism.

This paper is organized as follows.  We summarize the sample selection, photometry, and SFH measurement technique in \S \ref{sec:obs}.  In \S \ref{sec:results}, we explore empirical trends in the fraction of quenched galaxies as a function of mass and environment, and place our results in the context of the broader universe using data from the literature.  In \S \ref{sec:lgmodels}, we quantify delay times between infall and quenching by comparing our results with the model analysis presented in \citet{rocha2012}.  Finally, we summarize our findings in \S \ref{sec:summary}.   Throughout this paper, the conversion between age and redshift assumes the cosmology as detailed in \citet{planck2014xvi}. We recognize that several definitions of quenching epoch can be found in the literature.  For direct comparison, we have provided all SFH data in tabulated digital form in \citet{weisz2014a}.

\section{The Data}
\label{sec:obs}

\subsection{Galaxy Sample and Star Formation Histories}
\label{sec:sample}

Our sample contains 38 LG dwarf galaxies that all have deep archival HST/WFPC2 imaging.  The sample includes 25 diverse quenched galaxies (dwarf spheriodals, dSphs; and  dwarf ellipticals, dEs) spanning a large dynamic range in stellar mass (10$^4$ $\lesssim$ M$_{\star}$ $\lesssim$10$^8$ \msun).  Our sample also contains 5 transition dwarfs (dTrans), which have \hi\ but no evidence of on-going star formation \citep[i.e., no detectable \halpha; e.g.,][]{mateo1998}, and 8 star-forming dwarfs (dwarf irregulars, dIrrs), both of which provide good control and comparison samples.  The properties of our galaxy sample are listed in Table \ref{tab:alan_data}.

\begin{deluxetable*}{cccccccc}
\tablecaption{Observational Properties of our Local Group Dwarf Sample}
\tablecolumns{8}
\tablehead{
\colhead{Galaxy} &
\colhead{Morphological} &
\colhead{M$_{\mathrm{V}}$} &
\colhead{M$_{\star}$} &
\colhead{M$_{\mathrm{H{\sc I}}}$} &
\colhead{D$_{\mathrm{Host}}$} &
\colhead{r$_{\mathrm{h}}$}  &
\colhead{$\tau_{90}$}   \\
\colhead{Name} &
\colhead{Type} &
\colhead{} &
\colhead{(10$^{6}$ \msun)} &
\colhead{(10$^{6}$ \msun)} &
\colhead{(kpc)} &
\colhead{(pc)}  &
\colhead{log(yr)}   \\
\colhead{(1)} &
\colhead{(2)} &
\colhead{(3)} &
\colhead{(4)} &
\colhead{(5)} &
\colhead{(6)} &
\colhead{(7)}  &
\colhead{(8)}  
}
\startdata 
Andromeda {\sc I} & dSph & -11.7 & 3.9 & 0.0 & 58 & 672 &  9.86$^{+0.0,0.15}_{-0.05,0.14}$ \\
Andromeda {\sc II} & dSph & -12.4 & 7.6 & 0.0 & 184 & 1176 & 9.75$^{+0.01,0.07}_{-0.01,0.05}$ \\
Andromeda {\sc III} & dSph & -10.0 & 0.83 & 0.0 & 75 & 479 & 9.72$^{+0.05,0.3}_{-0.0,0.06}$  \\
Andromeda {\sc V} & dSph & -9.1 & 0.39 & 0.0 & 110 & 315 & 9.87$^{+0.02,0.23}_{-0.06,0.31}$ \\
Andromeda {\sc VI} & dSph & -11.3 & 2.8 & 0.0 & 269 & 524 & 9.73$^{+0.06,0.3}_{-0.02,0.09}$ \\
Andromeda {\sc XI} & dSph & -6.9 & 0.049 & 0.0 & 104 & 157 & 10.11$^{+0.0,0.0}_{-0.08,0.2}$ \\
Andromeda {\sc XII} & dSph & -6.4 & 0.031 & 0.0 & 133 & 304 & 9.56$^{+0.2,0.33}_{-0.04,0.04}$\\
Andromeda {\sc XIII} & dSph & -6.7 & 0.041 & 0.0 & 180 & 207 & 10.1$^{+0.0,0.0}_{-0.44,0.46}$ \\
Carina & dSph & -9.1 & 0.38 & 0.0 & 107 & 250 & 9.46$^{+0.0,0.11}_{-0.1,0.12}$ \\
Canes~Venatici {\sc I} & dSph & -8.6 & 0.23 & 0.0 & 218 & 564 & 9.92$^{+0.0,0.06}_{-0.1,0.11}$ \\
Canes~Venatici {\sc II} & dSph & -4.9 & 0.0079 & 0.0 & 161 & 74 & 9.92$^{+0.0,0.05}_{-0.24,0.24}$ \\
DDO~210 & dTrans & -10.6 & 1.6 & 4.1 & 1066 & 458 & 9.58$^{+0.07,0.52}_{-0.16,0.37}$\\
Draco & dSph & -8.8 & 0.29 & 0.0 & 76 & 221& 10.01$^{+0.01,0.06}_{-0.03,0.11}$ \\
Fornax & dSph & -13.4 & 20.0 & 0.0 & 149 & 710 & 9.38$^{+0.02,0.14}_{-0.02,0.06}$\\
Hercules & dSph & -6.6 & 0.037 & 0.0 & 126 & 330 & 10.1$^{+0.0,0.0}_{-0.61,0.61}$ \\
IC~10 & dIrr & -15.0 & 86.0 & 50.0 & 252 & 612 & 9.21$^{+0.02,0.15}_{-0.0,0.06}$\\
IC~1613 & dIrr & -15.2 & 100.0 & 65.0 & 520 & 1496 & 9.3$^{+0.0,0.08}_{-0.07,0.12}$ \\
Leo~A & dIrr & -12.1 & 6.0 & 11.0 & 803 & 499 & 8.78$^{+0.0,0.22}_{-0.0,0.08}$\\
Leo~{\sc I} & dSph & -12.0 & 5.5 & 0.0 & 258 & 251 & 9.23$^{+0.0,0.05}_{-0.0,0.0}$ \\
Leo~{\sc II} & dSph & -9.8 & 0.74 & 0.0 & 236 & 176 & 9.81$^{+0.0,0.05}_{-0.0,0.04}$ \\
Leo~{\sc IV} & dSph & -5.8 & 0.019 & 0.0 & 155 & 206 & 10.05$^{+0.0,0.04}_{-0.57,0.57}$ \\
Leo~T & dTrans & -8.0 & 0.14 & 0.28 & 422 & 120 & 9.23$^{+0.0,0.0}_{-0.02,0.05}$ \\
LGS~3 & dTrans & -10.1 & 0.96 & 0.38 & 269 & 470 & 9.58$^{+0.03,0.19}_{-0.02,0.09}$ \\
M32 & dE & -16.4 & 320.0 & 0.0 & 23 & 110 &  9.23$^{+0.0,0.49}_{-0.0,0.0}$ \\
NGC~147 & dE & -14.6 & 62.0 & 0.0 & 142 & 623 & 9.43$^{+0.02,0.53}_{-0.0,0.01}$\\
NGC~185 & dE & -14.8 & 68.0 & 0.11 & 187 & 42 & 9.56$^{+0.02,0.4}_{-0.02,0.05}$\\
NGC~205 & dE & -16.5 & 330.0 & 0.4 & 42 & 590 & 9.34$^{+0.0,0.47}_{-0.02,0.04}$\\
NGC~6822 & dIrr & -15.2 & 100.0 & 130.0 & 452 & 354 & 9.0$^{+0.04,0.58}_{-0.0,0.0}$ \\
PegDIG & dTrans & -12.2 & 6.61 & 5.9 & 474 & 562 & 9.23$^{+0.0,0.82}_{-0.11,0.11}$ \\
Phoenix & dTrans & -9.9 & 0.77 & 0.12 & 415 & 454 & 9.43$^{+0.02,0.08}_{-0.0,0.01}$\\
SagDIG & dIrr & -11.5 & 3.5 & 8.8 & 1059 & 282 & 8.91$^{+0.0,0.35}_{-0.06,0.3}$\\
Sagittarius & dSph & -13.5 & 21.0 & 0.0 & 18 & 2587 & 9.53$^{+0.0,0.19}_{-0.04,0.04}$\\
Sculptor & dSph & -11.1 & 2.3 & 0.0 & 89 & 283 & 10.03$^{+0.02,0.05}_{-0.01,0.17}$\\
Sex~A & dIrr & -14.3 & 44.0 & 77.0 & 1435 & 1029 & 8.85$^{+0.0,0.36}_{-0.0,0.07}$\\
Sex~B & dIrr & -14.4 & 52.0 & 51.0 & 1430 & 440 & 9.26$^{+0.02,0.79}_{-0.03,0.08}$\\
Tucana & dSph & -9.5 & 0.56 & 0.0 & 882 & 284 & 9.81$^{+0.0,0.2}_{-0.0,0.19}$\\
Ursa~Minor & dSph & -8.8 & 0.29 & 0.0 & 78 & 181 & 9.96$^{+0.01,0.07}_{-0.1,0.18}$\\
WLM & dIrr & -14.2 & 43.0 & 61.0 & 836 & 2111 & 8.85$^{+0.06,0.27}_{-0.0,0.0}$
\enddata
\tablecomments{Physical properties for galaxies in our sample taken from \citet{mcconnachie2012}.  The stellar mass listed in column (4) is computed from the integrated V-band luminosity and assumes \msun/\lsun $=$ 1.  Column (8) indicates the epoch at which 90\% of the stellar mass formed, including the random uncertainties (first error term) and the total uncertainty (second error term).  Both reflect their respective 68\% confidence intervals.  Note that for completeness we have also included these values for galaxies that are not currently quenched, i.e., dIrr/dTrans morphological types.} 
\label{tab:alan_data}
\end{deluxetable*}

We selected galaxies based on the availability of deep archival HST/WFPC2 observations.  All photometry was carried out using  \texttt{HSTPHOT}\footnote{\url{http://purcell.as.arizona.edu/hstphot/}}, a photometry package designed specifically for the under-sampled point spread function of WFPC2 \citep{dolphin2000a}, as part of the Local Group Stellar Populations Archive project\footnote{\url{http://astronomy.nmsu.edu/logphot}} \citep[LOGPHOT;][]{holtzman2006}.  To characterize completeness and observational biases in the photometry, we ran $\sim$ 10$^5$ artificial star tests per HST field, i.e., we inserted stars of a known magnitude into our images and recovered their photometry in an identical manner to the photometry of real stars.  Details of the photometric reduction process can be found in \citet{holtzman2006}.

All SFHs used in this paper were derived using the CMD fitting package \texttt{MATCH} \citep{dolphin2002}, as discussed extensively in \citet{weisz2014a}; we refer the reader to that paper for full details. Throughout the paper, we quote the 68\% confidence interval of the total uncertainties, including both random (i.e., statistical uncertainties) and systematic (i.e., a proxy for variations in the SFH as if they were solved with different stellar models) components.  Details of the uncertainty calculations for this sample are presented in \citet{weisz2014a}, and the general methodological details of the systematic and random uncertainty determinations are presented in \citet{dolphin2012} and \citet{dolphin2013}, respectively.

\subsection{Defining an Epoch of Quenching}
\label{sec:qepoch}

As a first step, we must define a quenching epoch for galaxies in our sample.  Identifying the quenching epoch for such a diverse collection of galaxies using CMD-based SFHs involves some subtlety, primarily owing to the presence of blue straggler stars (BSs).  BSs --- believed to be the result of merged or extreme mass transfer among low-mass main sequence stars \citep[e.g.,][and references therein]{davies2015} --- have similar optical colors and luminosities as intermediate age main sequence stars (2-5 Gyr old). Therefore, they may be mistaken as a signpost of intermediate age star formation, particularly in lower mass, quenched galaxies, where the `blue plume' is typically sparse.

Unfortunately, accounting for BSs in CMD-based SFH measurements is non-trivial.  Because there are no generative models for BSs, it is impossible to explicitly include them in the CMD fitting process.  One often-used solution is to simply exclude putative BSs from a CMD when measuring a SFH.  While this may be suitable for galaxies in which there is a strong prior belief that the `blue plume' is composed of only BSs \citep[e.g., the faintest MW satellites;][]{brown2012, brown2014}. It is challenging to implement this approach in a large, diverse sample of galaxies.  Several studies have shown that increasingly luminous galaxies have blue plumes that contain a larger fraction of genuine main sequence stars \citep[e.g.,][]{mapelli2007, momany2007, mapelli2009}.  Thus, excluding blue stars in \emph{all} early-type systems \textit{a priori} would mute the signatures of true intermediate age star-formation. Alternatively, excluding BSs in some galaxies and not in others is equally challenging, due to ambiguities in the point at which the `blue plume' is no longer purely BSs. Further, such an approach would compromise the homogeneity of our analysis.  Thus, as discussed in \citet{weisz2014a}, we have elected to not account for blue stragglers in the CMD modeling process, and instead rely on judicious interpretation of the resulting SFHs.

\begin{figure}
\begin{center}
\epsscale{1.2}
\plotone{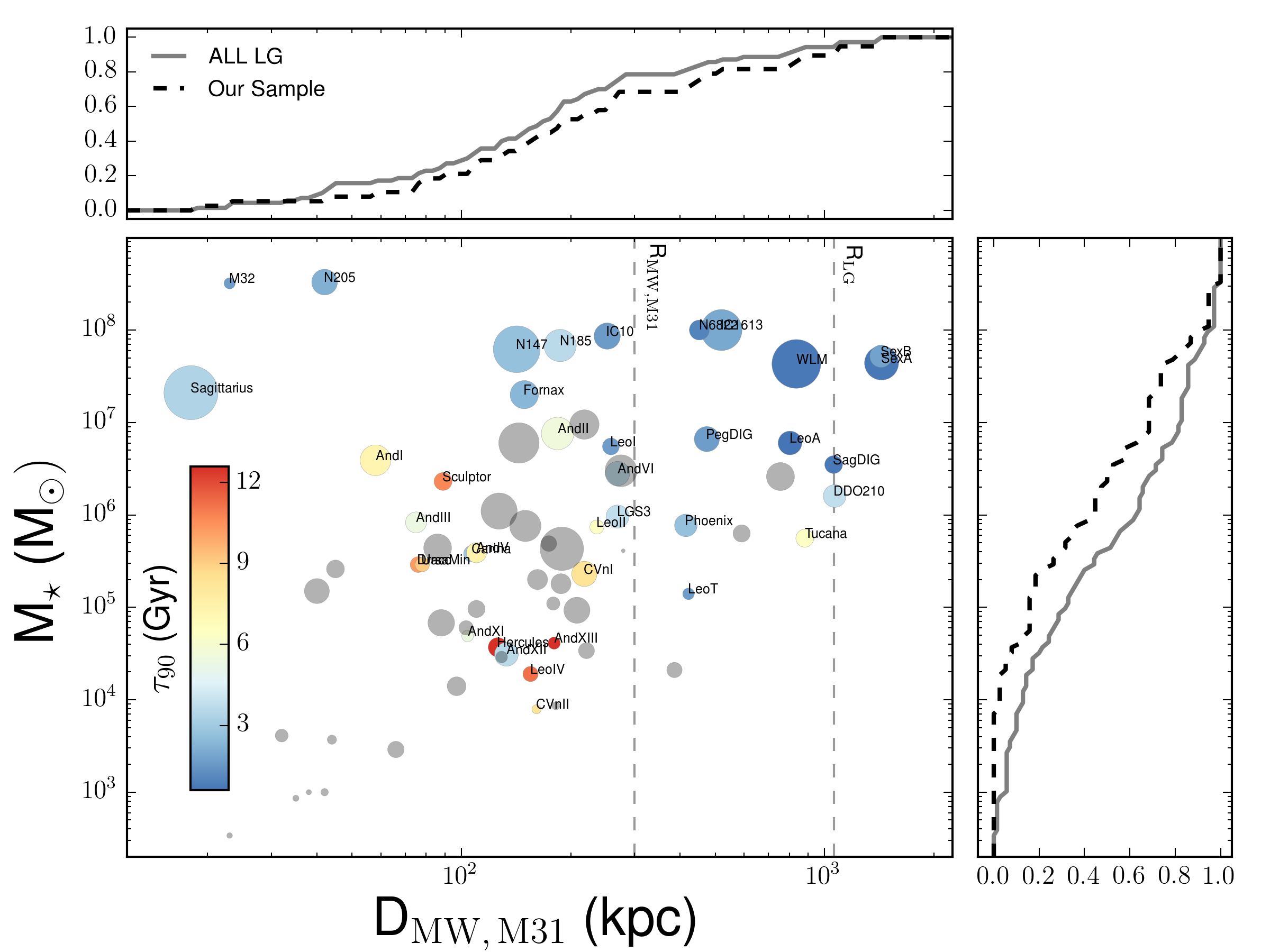}
\caption{Distance to the nearest host plotted versus present day stellar mass.  Galaxies in our sample are color-coded by \tauninety, the age at which 90\% of the total stellar mass formed.  Galaxies not in our sample are indicated in grey.  Point sizes are proportional to the half-light radii.  As shown by the cumulative distributions in the side panels, our sample provides reasonable representation of all LG dwarfs in terms of proximity to a massive host and present day stellar mass.  We are primarily missing the lowest mass systems located around the MW and the majority of the M31 companions, but do have some representation in both regimes.  In terms of \tauninety, there are broad trends such as lower mass galaxies typically have large values of \tauninety\ and more distant galaxies tend to not be quenched.  We discuss these trends in more detail in \S \ref{sec:results}.}
\label{fig:distmass_tau90}
\end{center}
\end{figure}

For the purpose of this paper, we use the lookback time at which a galaxy formed 90\% of its total stellar mass (hereafter \tauninety) as a proxy for the epoch at which a galaxy quenched.  This value was selected to minimize the impact of BSs on determining the quenching epochs of the least luminous systems in our sample.  The presence of BSs tends to shift the youngest \textit{bona fide} star formation to younger ages, while keeping about the same total mass formed in the system.  To estimate the impact of BSs on the SFHs, we re-ran SFHs of several lower luminosity dSphs having masked out the `blue plume' and found a typical mass contribution from BSs of $\sim$ 5\%, with an upper limit of $\lesssim$ 10\%.  Thus, while our adoption of \tauninety\ is a conservative choice, it  does avoid potential confusion of the quenching epoch due to the presence of BSs.  Readers who wish to explore other quenching metrics can electronically obtain all tabulated SFH data from \citet{weisz2014a}.

We list the values of \tauninety\ for each galaxy in Table \ref{tab:alan_data}.  These values were computed by interpolating the cumulative SFHs in linear time for each galaxy, including both random and systematic uncertainties, onto a fine time grid of 10 Myr in spacing.  This was done to pick out the exact 90\% value for each galaxy and avoid abrupt changes in the cumulative SFH that would compromise homogeneity in computing \tauninety\ from the native SFHs. For example, suppose the SFH abruptly jumps from 85\% in one bin to 98\% in a younger bin with a lull in star-formation in between.  Without interpolation we would be forced to pick the value closest to 90\%, whereas the interpolation allows to consistently select a value of \tauninety\ across all galaxies.  The assumption underlying the interpolation is that the SFH is constant across each bin.  However, the error introduced by this assumption is small compared to other sources of uncertainty (e.g., systematics).  

We plot \tauninety\ as a function of mass and distance from the nearest central galaxy in Figure \ref{fig:distmass_tau90}.  This plot shows many of well-known trends: currently star-forming galaxies are located at large distances from the MW or M31 \citep[e.g.,][]{grcevich2009}, lower mass systems tend to be preferentially quenched compared to higher mass systems \citep[e.g.,][]{tolstoy2009}. However, there appears to be interesting scatter in both trends, particularly when the values of \tauninety\ are considered.  We discuss these empirical relationships in more detail in \S \ref{sec:results}.

\subsection{Sample Completeness}
\label{sec:selection}

Sample completeness plays an important role in assessing quenching in low-mass systems.  Our sample is neither complete nor unbiased, as it was assembled based on the availability of HST/WFPC2 archival imaging.  However, as shown in Figure \ref{fig:distmass_tau90}, the sample is reasonably representative of the LG dwarf population in terms of stellar mass and proximity to a massive host. Specifically, our sample is $>$ 80\% complete for systems with M$_{\star}$ $\gtrsim$ 10$^6$ \msun, but only 10-40\% complete for galaxies with 10$^3$ $\lesssim$ M$_{\star}$ $\lesssim$ 10$^6$ \msun.  Similarly, our sample contains $\sim$ 50\% of the galaxies known to exist within the virial radius of the MW or M31, and 80\% of those located in the field.  Thus, while our sample has variable completeness in both mass and proximity to a host, it spans the entire range of properties of the LG dwarf population (dashed lines on the side panels).  In instances where completeness may compete with physical interpretation of the data we qualitatively discuss how it may affect our results.

\section{Time Evolution of the Quenched Galaxy Fraction}
\label{sec:results}

Having defined a proxy for the quenching epoch, we now examine the relationship between \tauninety\ and physical properties, such as stellar mass and proximity to a massive host.

\subsection{Quenched Fraction as a Function of Stellar Mass and Lookback Time}
\label{sec:quenchedmass}

\begin{figure}
\begin{center}
\epsscale{1.2}
\plotone{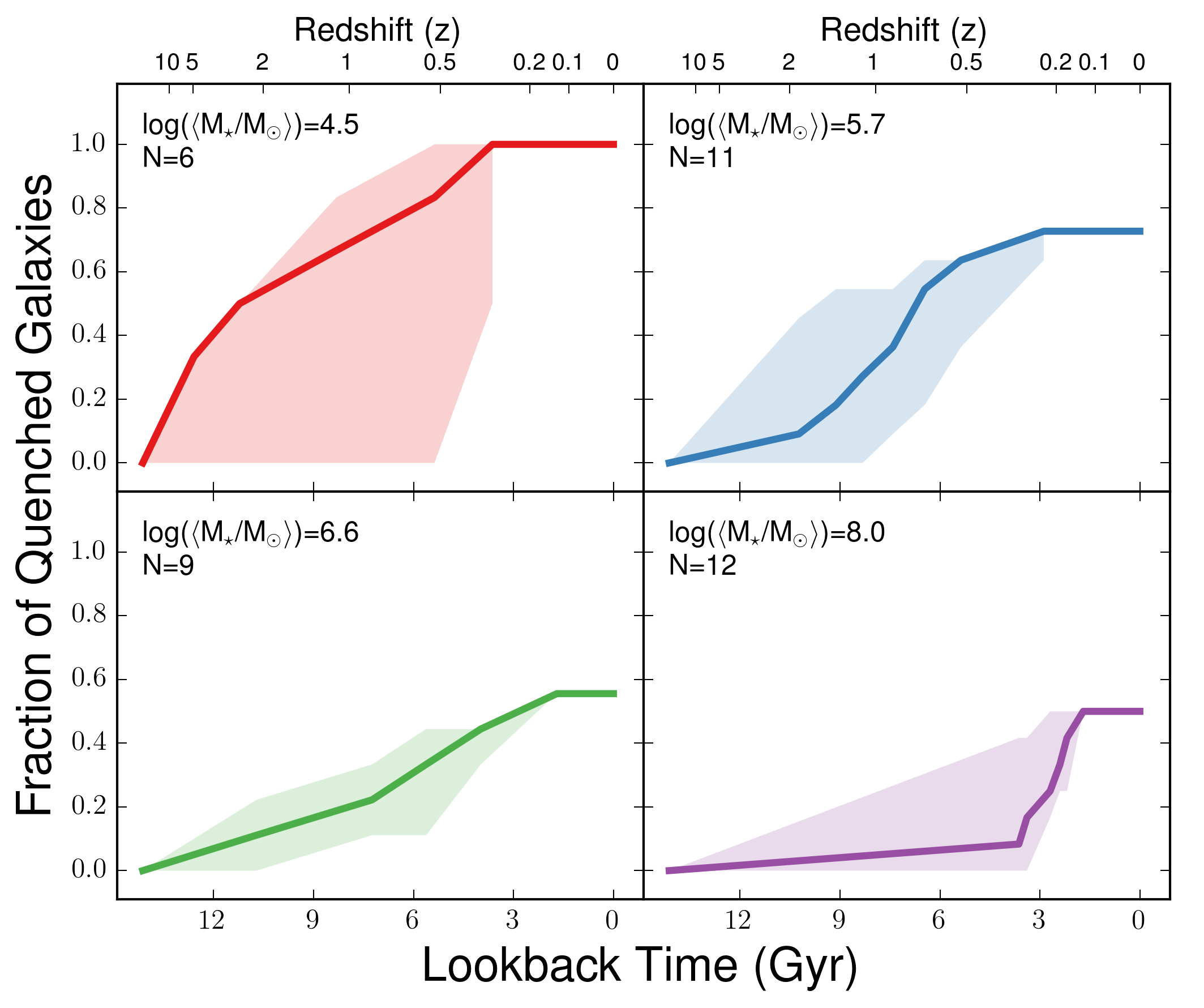}
\caption{The quenched fraction of LG dwarfs plotted as a function of present day stellar mass and lookback time. The solid lines trace the most likely quenched fraction values, while the uncertainty envelopes represent the the 68\% confidence interval in the quenched fraction, including both systematic and random uncertainties.  The best fits show a broad trend in quenched fraction as a function of mass.  The quenched fraction is higher for lower mass galaxies at nearly all epochs considered.  Further, as a population, based on the quenched fraction, higher mass galaxies tend to have quenched later compared to lower mass galaxies.}
\label{fig:mass_tau90}
\end{center}
\end{figure}

In Figure \ref{fig:mass_tau90}, we plot the fraction of quenched galaxies as a function of lookback time in four equally spaced logarithmic bins of present day stellar mass (log(M$_{\star}$/\msun)$=$4.0-5.0, 5.0-6.0, 6.0-7.0,  $>$7.0; hereafter we refer to these mass bins by their unweighted mean masses).  We have defined the quenched fraction to be the fraction of galaxies that quenched at earlier an epoch than a specified lookback time (i.e., \tauninety $\ge$ $t$).  The shaded envelopes reflect the narrowest 68\% confidence interval of the total uncertainty (i.e., random and systematic, which are inclusive of Poisson variance) that includes the best fit value of \tauninety\ (traced by the solid line).  These uncertainty estimates are slightly conservative (i.e., too large) as they include possible co-variances in the SFHs that were derived from CMDs of similar depths, as discussed in \citet{weisz2014a}.

Figure \ref{fig:mass_tau90} shows a clear trend for lower mass dwarfs: they have a larger quenched fraction than more massive dwarfs at every epoch.  By $\sim$10 Gyr ago, $\sim$50\% of the lowest mass systems (\logavgmass = 4.5) were quenched, which is at least a factor of 5 higher than the quenched fraction in the other three mass ranges considered. The quenched fraction in the lowest mass bin steadily grows until $\sim$4-5 Gyr ago, by which time the quenched fraction has reached unity.

In the second lowest mass bin (\logavgmass = 5.7), the quenched fraction grows steadily from $\sim$ 10\% at $\sim$ 10 Gyr ago to $\sim$ 70\% at $\sim$ 3 Gyr ago.  At higher masses, the quenched fraction is even lower.   In the second highest mass bin (\logavgmass = 6.6), the quenched fraction reaches 50\% at $\sim$ 2.5 Gyr ago.  In the highest mass bin (\logavgmass = 8.0), the quenched fraction at the present is only $\sim$ 40\%.  

The temporal evolution of the quenched fraction can also provide clues as to how quickly dwarf galaxies can undergo quenching. For example, in the lowest mass systems, nearly 40\% appear to be quenched by $\sim$ 12 Gyr ago, which implies a quenching timescale of $\lesssim$ 1.5-2 Gyr.  Such a timescale is consistent with quenching due to reionization or early-time environmental processes.  Interestingly, the quenched fraction of the lowest mass galaxies continues to grow toward the present, indicating that not all low-mass systems were equally affected by quenching mechanisms that operated exclusively in the early universe (e.g., reionization).  In contrast, galaxies in all other mass ranges have low quenched fractions at comparably early epochs. This broad trend is consistent with predictions from some reionization models, in which only the lowest mass systems were significantly affected by reionization \citep[e.g.,][]{ricotti2005}.  However, other models suggest that the influence of reionization on low-mass galaxies is more complex than permanently quenching star-formation $\sim$ 12 Gyr ago \citep[e.g.,][]{benitezllambay2014, onorbe2015}.  We refer the reader to \citet{weisz2014b} for an in-depth discussion of signatures reionization in the SFHs of low-mass galaxies.

In the two intermediate mass ranges, the nearly uniform increase in quenched fraction toward the present provides little information about the quenching timescale. In contrast, the highest mass galaxies are more informative for about the rapidity of quenching.  In this mass range (\logavgmass = 8.0), the quenched fraction grew from 0 to 40\% between 2 and 4 Gyr ago.  This rapid rise indicates that galaxies of this mass \emph{can} quench on fairly rapid timescales, although without a knowledge of the infall times we cannot determine the exact quenching timescale (see \S \ref{sec:lgmodels}).  Further, the six systems in this mass range may not be broadly representative of galaxies in this mass range, given that four are dEs in M31 group (M32, NGC~147, NGC~185, NGC~205) and the other two are MW companions Fornax and Sagittarius.  The nearly synchronized quenching of M31 dEs may be indicative of a global event in the M31 group \citep[e.g., simultaneous satellite accretion, major merger;][]{fardal2008, hammer2010}, which may detract from the generality of this finding.  For a further discussion of quenching in M31 satellites, we refer the reader to \S \ref{sec:mwm31} and \citet{weisz2014m31}.

Up to this point, we have not considered the role of uncertainties and selection effects in our results and interpretation. The uncertainties on the quenched fraction are most noticeable for the lowest mass galaxies, which typically show broad 68\% confidence intervals on their values of \tauninety.  This breadth is the results of a small number of stars on the CMDs of the lowest mass systems, as well as the presence of BSs, which introduce a large dispersion into \tauninety. For the more massive systems, which have more populated CMDs, the effects of uncertainties in the quenched fraction are noticeably less pronounced.

Completeness of the sample can also affect our findings.  Many of the faint MW galaxies that are not in our sample appear to have quenched at lookback times of $\gtrsim$ 10-12 Gyr ago \citep[e.g.,][]{dejong2008a, sand2009, sand2010, sand2012, brown2012, brown2014, okamoto2012}.  A naive completeness correction, in which we assume that all missing galaxies are similar to the faint MW companions, implies that the quenched fraction could be as high as $\sim$80\% by 10-12 Gyr ago. However, as demonstrated in \citet{weisz2014m31}, not all quenched low-mass systems in the LG stopped forming stars $>$ 10 Gyr ago.  Deep imaging of Andromeda {\sc XVI} (M$_{\star}$ $\sim$ 10$^5$ \msun) reveals a quenching epoch of $\sim$ 5 Gyr ago. Similarly, other low-mass galaxies in our sample such as Andromeda {\sc XII} and Leo~T have significant amounts of star formation $<$ 10 Gyr ago, even when full uncertainties are considered.  Thus, on the whole, the simple assumption that all of the lowest-mass quenched galaxies are similar to the faintest MW satellites may not be an accurate extrapolation.  Finally, we note our sample is 80\% complete for M$_{\star} \gtrsim$ 10$^6$ \msun, indicating that incompleteness in this regime is not as important as for lower mass systems.

\subsection{Quenched Fraction as a Function of Environment}
\label{sec:quenchedenvironment}

\begin{figure}
\begin{center}
\epsscale{1.2}
\plotone{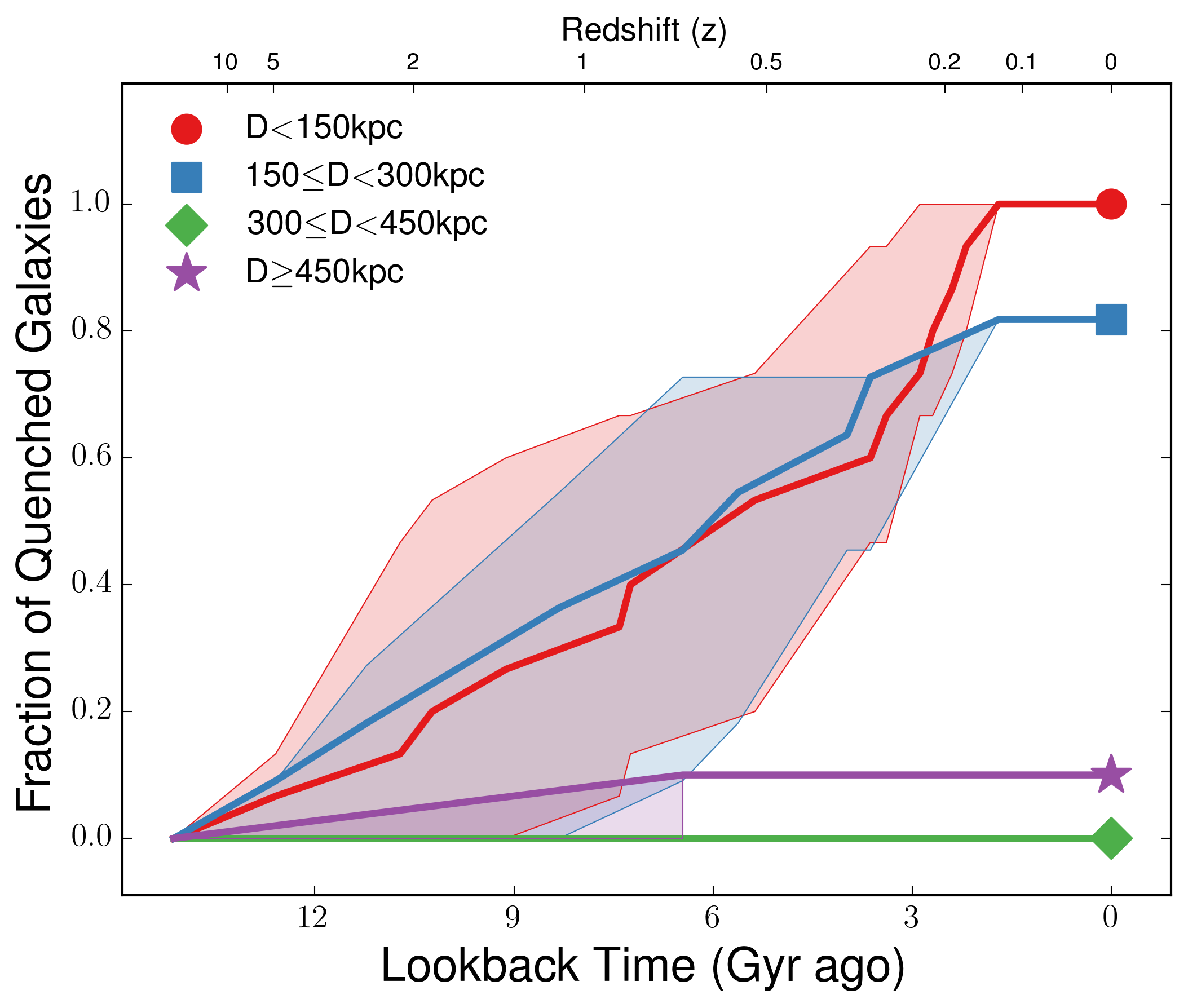}
\caption{The same as Figure \ref{fig:mass_tau90}, except for proximity to M31 or the MW instead of stellar mass.  Galaxies inside the virial radius of M31 or the MW are largely quenched (the exceptions are IC~10 and LGS~3), while those outside the virial radius are not (akin to the LG morphology-density relationship).  However, within 300 kpc there is no clear trend between present day distance from a host and quenching epoch.}
\label{fig:dist_tau90}
\end{center}
\end{figure}

In Figure \ref{fig:dist_tau90}, we plot the quenched fraction as a function of lookback time in bins of current distance from a massive host.  We have divided the sample into four equally spaced radial bins in multiplies of \rvir, ranging from R$<$0.5 \rvir\ to R$>$1.5 \rvir, where we adopt \rvir$=$300 kpc for both M31 and the MW \citep{mcconnachie2012}.   

From this plot, it is clear that environment plays an important role in quenching low-mass galaxies.  At the present day, few dwarfs located outside \rvir\ are quenched, while those located inside \rvir\ are almost entirely quenched. This trend is essentially a re-statement of the long-established LG morphology-density relationship \citep[e.g.,][and references therein]{mateo1998, vandenbergh2000, grcevich2009}.  

For dwarfs located inside \rvir, the temporal evolution of the quenched fraction is essentially independent of the current proximity to M31 or the MW, and is statistically indistinguishable in both bins within R$<$300 kpc. The small differences at recent times are due to the presence of gas-rich galaxies LGS~3 and IC~10 inside the virial radius of M31.  The independence between quenching time and current proximity to a massive host is consistent with a picture in which quenching occurs when or soon after a galaxy enters the virial radius.  This conclusion does not depend on selection effects and uncertainties in the values of \tauninety.

In principle, the SFHs may also provide clues to the planar satellite structures that are known to exist around the MW and M31 \citep[e.g.,][]{lyndenbell1976, kroupa2005, ibata2013}.  For example, if the planes were accreted as a group at the same time, the process of accretion may leave systematic signatures in the quenching epochs that are different from off-plane systems.  Unfortunately, we do not detect any such clear trends.  In the case of M31, only two of our galaxies, And {\sc II} and And {\sc V} are off-plane, limiting any general comparison.  Similarly for the MW sample, there is no clear difference in the SFHs of planar and non-planar satellites.  The lack of a correlation does not rule out any large scale accretion scenario, as dynamical mixing may have erased the initial conditions of any such event.

\subsection{Quenching Properties of the Milky Way and M31 Satellites}
\label{sec:mwm31}

\begin{figure}
\begin{center}
\epsscale{1.2}
\plotone{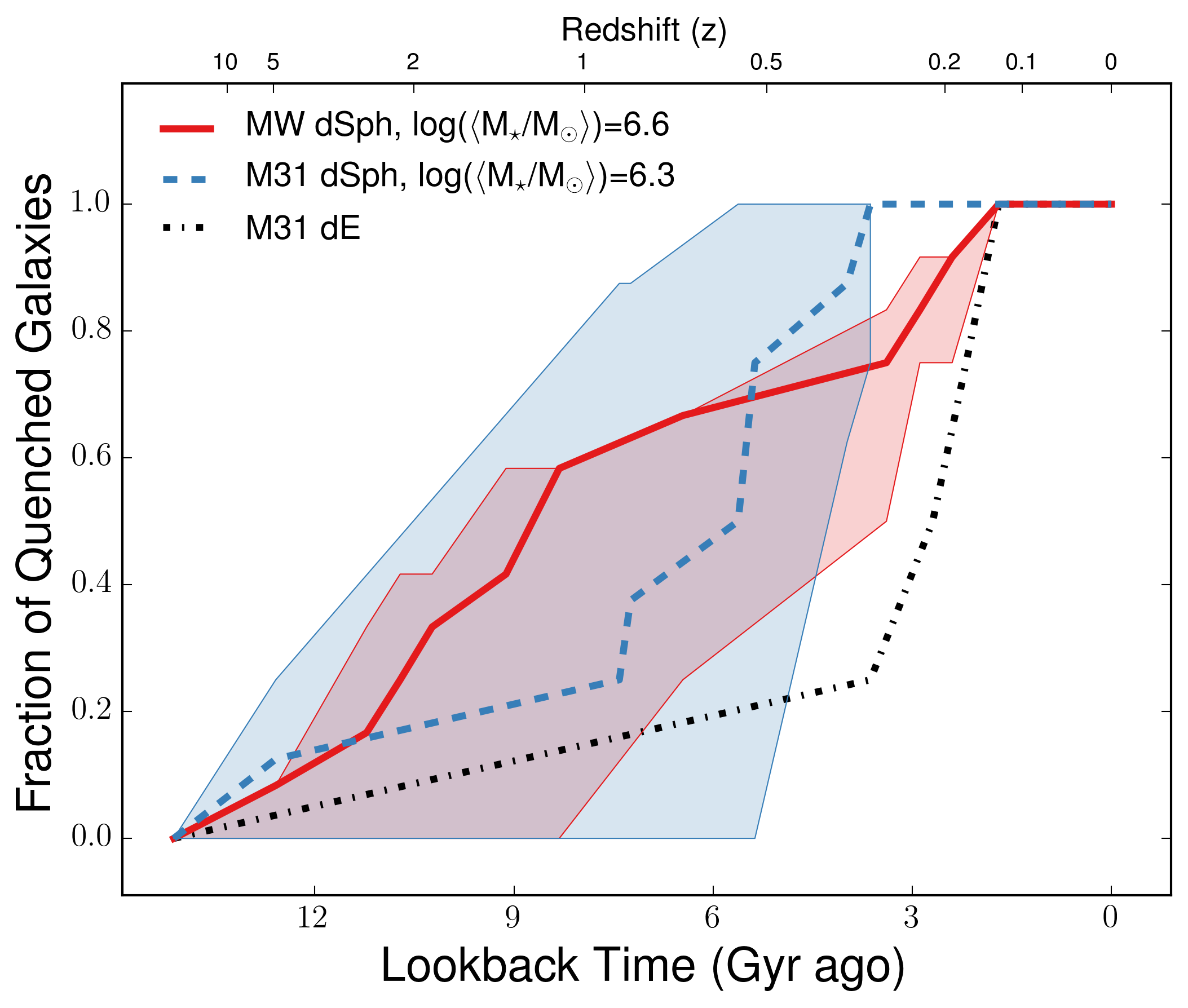}
\caption{Time evolution of the quenched fraction of MW dSphs (red), M31 dSphs (blue), and M31 dEs (dashed line; shown only for qualitative reference).  The most likely quenched fractions (points) appear different for the M31 and MW dSphs.  The quenched fraction of M31 dSphs increased rapidly at intermediate ages (5-7 Gyr ago), while the quenched fraction of MW dSphs has steadily grown over the history of the universe.  These contrasting pattens may provide clues to differences in the evolution of the sub-groups, however the uncertainties are too large to draw definitive conclusions.  In comparison to the dSphs, the quenched fraction of the M31 dEs increased rapidly between 2 and 3 Gyr ago.}
\label{fig:mwm31_tau90}
\end{center}
\end{figure}

As discussed in \S \ref{sec:quenchedmass}, there are hints that the M31 and MW satellites may have different quenching characteristics. We compare these two systems in detail in Figure \ref{fig:mwm31_tau90}.  Here, the quenched fractions show qualitatively different trends.  While the quenched fraction for the MW satellites steadily increases toward the present, the quenched fraction for M31 dSphs jumps from 20\% at $\sim$ 7 Gyr ago to 100\% by $\sim$ 5 Gyr ago.  Further, the M31 dEs appear to have rapidly quenched between $\sim$ 2 and 4 Gyr ago, a different timescale than both sets of dSphs.  

Taken at face value, these results suggest that satellites in the two sub-groups may have systematically different quenching characteristics, despite the MW and M31 having similar stellar masses.  The broader implication is that satellite behavior may be sensitive to the specific accretion history of the host galaxy, as opposed to simply scaling with host mass.  However, our data only provide hints of these differences due to large uncertainties in the SFHs of M31 satellites and incompleteness in both the MW and M31 satellite samples.  Indeed the uncertainties allow there to be identical quenching timescales between the two systems. Upcoming HST programs aimed at systematically imaging the M31 satellites (GO-13739, PI:Skillman; GO-13699, PI: Martin) should provide the data necessary to make more definitive statements about the quenching epochs of M31 satellites \citep[e.g.][]{weisz2014m31}.

\subsection{The Quenched Fraction over 7 dex in Stellar Mass}
\label{sec:ushape}

The fraction of quenched galaxies as a function of redshift and stellar mass is widely used as an observational constraint for modeling quenching processes in groups and clusters of galaxies \citep[e.g.,][]{mandelbaum2006, vandenbosch2008, peng2010, bauer2013, moustakas2013,  wetzel2013}.  However, the majority of such studies have been limited to galaxies with M$_{\star}$ $\gtrsim$ 10$^9 - 10^{10}$ \msun, since fainter galaxies are extremely challenging to detect beyond the very nearby universe.  As a result, there is no unified picture of quenching properties and processes over a full compliment of galaxy masses and lookback times.

At the same time, the observational evidence that quenching in lower mass galaxies is strongly driven by environment is steadily growing.  For example, \citet{geha2012} show the fraction of quenched galaxies with M$_{\star}$ $\sim$ 10$^{8.5}$ \msun\ are rare at distances $>$ 1.5 Mpc from a massive host.  Within the nearest few Mpc, KKR~25 and KKs3 are only known examples of quenched, isolated dwarf galaxy \citep[e.g,][]{weisz2011a, makarov2012, karachentsev2015}.  The presumed\footnote{It is not possible to rule out completeness effects in the current low-mass galaxy surveys. However, any attempts to quantify the numbers and characteristics of undiscovered low-mass field galaxies would be pure speculation.  We therefore take the current results at face value while acknowledging the competing effects of completeness.}  The rarity of quenched, low-mass field galaxies suggests that most low-mass galaxies quenched in higher-density environments, although it is not possible to rule out completeness effects in the current surveys. 

\begin{figure}
\begin{center}
\epsscale{1.2}
\plotone{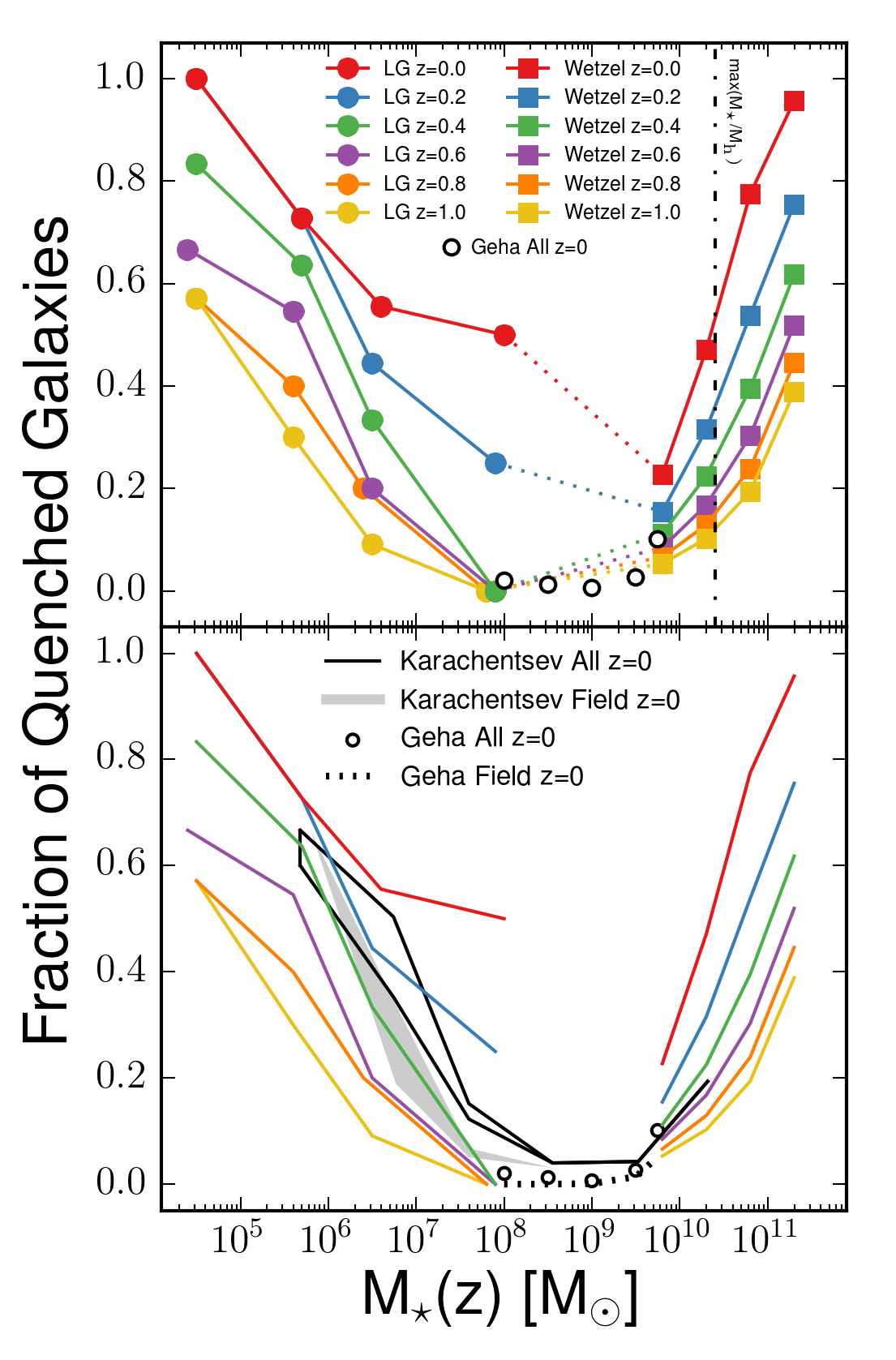}
\caption{\small The evolution of the quenched fraction of all galaxies over $\sim$ 7 dex in stellar mass from $z=0-1$. {\textit Top--}The quenched fraction for low mass galaxies (M$_{\star}$ $\lesssim$ 10$^8$ \msun) are from this paper, and the higher mass systems (M$_{\star}$ $\gtrsim$ 10$^8$ \msun) are from \citet{geha2012} and \citet{wetzel2013}.  {\textit Bottom --}  The thin colored lines are the same as in the top panel, while the thick lines are the quenched fraction of all and field galaxies from \citet{geha2012} and \citet{karachentsev2013}, as described in \S \ref{sec:ushape}.  The envelopes for the \citet{karachentsev2013} data shows the range of quenched fractions for whether dTrans are considered quenched or not. In both panels, the ``U'' shape in the the quenched fraction as a function of mass may reflect the mass scales at which different quenching mechanism operate.  At low masses, external processes (e.g., reionization, ram pressure, tidal stripping) are thought to drive quenching, while quenching in higher mass systems is controlled by halo mass (e.g., gas accretion rates, AGN feedback, virial shocks, stellar feedback).  The ubiquitously low quenched fraction at intermediate masses (M$_{\star}$ $\gtrsim$ 10$^8$-10$^{10}$ \msun) may indicate a transition region in which neither type of quenching mechanism is particularly effective.}
\label{fig:quench_zred}
\end{center}
\end{figure}

Therefore, exploring a broader picture of quenching requires comparing galaxies with respect to both mass and environment.  In the remainder of the section, we compared quenching in the LG dwarfs with higher mass galaxies from all environments.  In the next section, we discuss the LG dwarfs alongside satellite populations to assess the specific role of environment.  

In Figure \ref{fig:quench_zred} and Table \ref{tab:lgquench}, we illustrate the evolution of the quenched fraction from $z=0-1$ ($\sim$ 0-8 Gyr ago) for galaxies spanning a range of 7 dex in stellar mass.  In the top panel, we have combined the quenched fractions derived in this paper for galaxies with M$_{\star}$ $\lesssim$ 10$^8$ \msun\ from \citet{wetzel2013}  (M$_{\star}$ $\sim$ 10$^{10}$ $\lesssim$ M$_{\star}$ $\lesssim$ 10$^{11.5}$ \msun; $z=0-1$). This figure provides a first view of the quenched fraction over large ranges in stellar mass and redshift.  Note that for consistency with the analysis of \citet{wetzel2013}, we compute the LG dwarf stellar masses at the specified redshift.  There are generally negligibly different than the $z=0$ masses used elsewhere in the paper, as most dwarfs only formed a small fraction of their mass at $z\le 1$. 

The most notable feature in this plot is the minimum of the quenched fraction as a function of mass, at a given redshift.  At each redshift considered, the quenched fraction steadily declines from maxima at the highest and lowest stellar masses toward a minimum at M$_{\star}$$\sim$10$^8$-10$^{10}$ \msun.   The ``U" shape in the  fraction of quenched galaxies may provide clues for which quenching mechanisms operate at different mass scales, which we now discuss in more detail.

At the highest masses (M$_{\star}$$\gtrsim$10$^{11}$ \msun), quenching may be driven by several internal processes including AGN feedback, virial shock heating, restricted gas accretion, stellar feedback and/or disk instabilities \citep[e.g.,][]{silk1998, keres2005, birnboim2007, dekel2009, woo2013}.  Although the relative contribution of each mechanism to quenching remains unclear, the broad picture is that the depth of the potential well (i.e., halo mass) largely drives the initial quenching by depleting gas and suppressing cooling.  Further, star formation may be prevented by a combination of mechanisms (such as those above) or via other pathways such as heating from asymptotic giant branch stars \citep[e.g.,][]{conroy2014} that appear incapable of initial quenching, but may provide sufficient heating to maintain a massive galaxy's quenched state.

\begin{deluxetable}{ccccc}
\tablecaption{Quenched Fraction of Galaxies as a Function of Redshift and Mass}
\tablecolumns{5}
\tablehead{
\colhead{log(M$_{\star}(z)$)} &
\colhead{Quenched} &
\colhead{} &
\colhead{log(M$_{\star}(z)$)} &
\colhead{Quenched} \\
\colhead{(M$_{\odot}$)} &
\colhead{Fraction} &
\colhead{} &
\colhead{[(M$_{\odot}$)} &
\colhead{Fraction} \\
\colhead{(1)} &
\colhead{(2)} &
\colhead{(3)} &
\colhead{(4)} &
\colhead{(5)} 
}
\startdata 
$z=0$  & & & $z=0.2$  &  \\
\hline
4.5 & 1.00 & & 4.5 &  1.0 \\
5.7 & 0.73 & & 5.7 &  0.73 \\
6.6 & 0.56 & & 6.5 &  0.44 \\
8.0 & 0.50 & & 7.9 & 0.25 \\
\hline
$z=0.4$  & & & $z=0.6$   & \\ 
\hline
4.5 & 0.83 & & 4.5 &  0.67 \\
5.7 & 0.64 & & 5.5 &  0.55 \\
6.5 & 0.33 & & 6.5 &  0.2 \\
7.9 & 0.00 & & 7.9 & 0.00 \\
\hline
$z=0.8$  & & & $z=1.0$  &  \\
\hline
4.5 & 0.57 & & 4.5 &  0.57 \\
5.6 & 0.40 & & 5.6 &  0.3 \\
6.4 & 0.2 & & 6.4 &  0.1 \\
7.8 & 0.00 & & 7.8 & 0.00
\enddata
\tablecomments{The fraction of quenched LG dwarf galaxies as a function of stellar mass and redshift.  This table corresponds to the values plotted in Figures \ref{fig:quench_zred} and \ref{fig:quench_satellite}.  We emphasize that due to selection effects, these values should be considered in a qualitative sense, as discussed in \S \ref{sec:ushape}. }
\label{tab:lgquench}
\end{deluxetable}

At slightly lower masses (M$_{\star}$$\sim$10$^{10}$ \msun), halo mass may play a diminished role in quenching.  The ratio of stellar-to-halo mass reaches a global maximum at M$_{\star}$$\sim$10$^{10.5}$ \msun \citep[e.g.,][]{moster2010, behroozi2013}, suggesting that stellar feedback may play a more important role in quenching in this mass range \citep[e.g.,][]{moster2010, hopkins2011, hopkins2012, hopkins2014}.

\begin{deluxetable*}{cccccc}
\tablecaption{Quenched Fraction For the Nearby Galaxy Catalog \citet{karachentsev2013}}
\tablecolumns{6}
\tablehead{
\colhead{$\langle$log(M$_{\star}$/M$_{\odot}$)$\rangle$} &
\colhead{Quenched Fraction} &
\colhead{N$_{\rm Total}$} &
\colhead{N$_{\rm Quenched}$} &
\colhead{Quenched Fraction} &
\colhead{N$_{\rm Quenched}$} \\
\colhead{} &
\colhead{} &
\colhead{} &
\colhead{dTrans$=$} &
\colhead{} &
\colhead{dTrans $=$} \\
\colhead{} &
\colhead{} &
\colhead{} &
\colhead{star-forming} &
\colhead{} &
\colhead{not star-forming} \\
\colhead{(1)} &
\colhead{(2)} &
\colhead{(3)} &
\colhead{(4)} &
\colhead{(5)} &
\colhead{(6)}
}
\startdata 
& & & All Galaxies  & &    \\ 
\hline
5.68 & 0.60 & 15 & 9  & 0.67 & 10 \\
6.74 & 0.35 & 139 & 49  & 0.50 & 70\\
7.60 & 0.13 & 310 & 38  & 0.15 & 47 \\
8.55 & 0.04 & 201 & 8  & 0.04 & 8  \\
9.53 & 0.04 & 94 & 4  & 0.04 & 4\\ 
10.32 & 0.19 & 31& 6 & 0.19 & 6 \\
\hline
& & & Field Galaxies  & &   \\
\hline
5.80 & 0.67 & 6 & 4   & 0.67 & 4\\
6.77 & 0.19 & 74 & 14   & 0.33 & 25\\
7.60 & 0.05 & 237 & 12   & 0.07 & 16 \\
8.55 & 0.03 & 164 & 5 & 0.03 & 5 \\
9.51 & 0.01 & 80 & 1  & 0.01 & 1\\ 
\hline
& & & Satellite Galaxies  & &  \\
\hline
5.58 & 0.56 & 9 & 5  & 0.67 & 6\\
6.71 & 0.54 & 65 & 35  & 0.69 & 45\\
7.60 & 0.36 & 72 & 26  & 0.43 & 31 \\
8.53 & 0.08 & 37 & 3 & 0.08 & 3 \\
9.60 & 0.21 & 14 & 3  & 0.01 & 3

\enddata
\tablecomments{The $z=0$ quenched fraction of galaxies as determined from analysis of the \citet{karachentsev2013} nearby galaxies catalog (D $<$ 11 Mpc).  The quenched fractions were computed by comparison of listed morphological types and 3D distances as described in \S \ref{sec:ushape}.  Due to potential confusion between dSph and dTrans morphological types at low stellar masses, we have included quenched fractions in which dTrans are counted star-forming (columns (4) and (5)) or are counted as quenched (columns (5) and (6)).  LG galaxies have been exulted from this analysis.}
\label{tab:kara_data}
\end{deluxetable*}

The next lowest mass data points (M$_{\star}$$\lesssim$10$^8$ \msun) are from the LG sample.   Compared to the \citet{wetzel2013} data, these galaxies are (a) highly dark matter dominated and (b) represent an obvious transition to a group environment.  As a result, qualitatively different quenching mechanisms may be at work.  Typically, quenching of galaxies in this mass range is ascribed to environmental processes such as ram pressure or tidal stripping \citep[e.g.,][]{mayer2001, mayer2006}, both of which require the presence of a massive galaxy, i.e., a group environment, to be effective. Putative environmentally independent quenching mechanisms such as stellar feedback and gas-consumption due to star formation \citep[e.g.,][]{dekel1986}, are not highly viable quenching mechanisms.  Stellar feedback creates outflows of hot gas, but is not energetic enough to completely unbind cold gas from dark matter halos \citep[e.g.,][]{maclow1999, governato2010, governato2012}.  Similarly, gas consumption from star formation is not a likely quenching mechanism.  Gas consumption timescales for most gas-rich dwarfs are tens to hundreds of Gyr \citep[e.g.,][]{vanzee1997, skillman2003a}, and, if acting alone, would produce few quenched dwarfs in a Hubble time.

At the lowest masses (M$_{\star}$$\lesssim$10$^6$ \msun), reionization may potentially playing a role in quenching \citep[e.g.][]{efstathiou1992, bullock2000, ricotti2002a}.  Although often treated as a global, and thus an environmentally independent process, detailed simulations suggest that reionization is not necessarily uniform in time, intensity, and/or environment; therefore its impact on low mass systems may be highly variable \citep[e.g.,][]{busha2010, ocvirk2013}.  As discussed in \citet{weisz2014b} and in \S \ref{sec:quenchedmass}, large variability in the SFHs of the lowest mass galaxies may be consistent with an inhomogenous reionization scenario.

Unfortunately, data from the LG and \citet{wetzel2013} provide no sampling in the minimum mass range (M$_{\star}$$\sim$10$^8$-10$^{10}$ \msun).  However, nearby galaxy surveys can fill in this range at $z=0$ providing some sampling for the behavior of these systems.  In the bottom panel of Figure \ref{fig:quench_zred} we add the quenched fractions for all and field galaxies from the SDSS-based study of \citet{geha2012} and from the nearby galaxy catalog of \citet{karachentsev2013}.  For the former dataset, field galaxies are those with projected distances $>$ 0.5 Mpc from a massive host.  

For the latter, we have estimated the quenched fraction from the nearby galaxy catalog (1$<$D$<$11 Mpc) of \citet{karachentsev2013} as follows.  We first convert all listed M$_{B}$ values into stellar masses assuming \msun/\lsun $=$1 and a M$_{B, \odot}$ $=$5.45.  We designate all galaxies with M$_{\star} >$ 10$^{10}$ \msun\ to be `central' galaxies, and compute the 3D distances (based on the listed line of sight distances and sky coordinates) between each central and the other 861 galaxies in the catalog.  Galaxies that are within 0.5 Mpc of a central are considered to be satellites, and more distant non-centrals are designated as field galaxies.  We then compute the quenched fraction by comparing listed morphologies; those with T$<$0 are considered to be quenched and those with T$\ge$0 are not.  At M$_{\star}$ $\lesssim$ 10$^7$ \msun, the \citet{karachentsev2013} suffers from decreasing completeness, and the authors also note that there is some ambiguity in morphological typing between star-forming transition objects that have faint GALEX fluxes but no detectable \halpha\ (i.e., dTrans) and gas-free, non-star forming dSphs.  To mitigate this latter effect, we have computed the morphological-based quenched fractions in both cases (i.e., all are dTrans are quenched or all dTrans are star-forming).  These ranges are reflected by the envelops in the bottom panel of Figure \ref{fig:quench_zred}. Throughout this process, all LG galaxies were excluded from consideration. As discussed below, the quenched fractions determined by this method may not be as quantitative as that of \citet{geha2012}, but this does provide an extension to lower masses that is not possible with SDSS. As a sanity check, the morphological approach provides very good overall agreement with expectations from both LG and ANGST SFHs.  We list the  quenched fractions derived from the \citet{karachentsev2013} in Table \ref{tab:kara_data}.

The addition of literature data at $z=0$ nicely fills in missing mass range from the top panel of Figure \ref{fig:quench_zred} and helps paint a picture of quenching trends over a wide range of stellar mass.  For the intermediate mass range (M$_{\star} \sim$ 10$^{8}$ - 10$^{10}$  \msun), the quenched fractions from \citet{geha2012} and \citet{karachentsev2013} are nearly zero.  Both exhibit a small to modest upturn at M$_{\star} \gtrsim$ 10$^{10}$ \msun, which is qualitatively consistent with the \citet{wetzel2013} results.  

At M$_{\star} \lesssim$ 10$^{8}$ \msun, the \citet{karachentsev2013} quenched fractions show a significant upturn.  This is particularly interesting for the field sample, as is it not clear what mechanisms can effectively quench such galaxies in the absence of environment.  One possibility is that there are so-called `backsplash' galaxies, which had an interaction with a massive host and have since been ejected outside of the virial radius, placing them into the quenched field galaxy category.  Similarly, analysis of the ELVIS Nbody simulations \citep{garrisonkimmel2013} by \citet{deason2014} suggest that dwarfs outside the virial radius of a massive galaxy have a moderately high merger rate, which may be capable of inducing quenching.  Alternatively, it may be the case that lower-mass systems are more vulnerable to gas-loss due to feedback than current simulations suggest, which would introduce a mass-dependent quenching trend in isolated systems.  Suggestions that dwarfs do not follow the stellar-halo mass relation from higher mass galaxies \citep[e.g.,][]{garrisonkimmel2014}, i.e., they more form more stars for a given halo mass, may also play a role in this possibility.  Taken at face value, these data indicate that there are a modest number of quenched low-mass galaxies in the nearby universe, although there does not remain a clear physical reason for the observed quenched fraction in the field particularly as the majority of galaxies in this sample are too massive to have been completely quenched by photo-heating due to cosmic reionization \citep[e.g.,][]{bullock2000, ricotti2005}.

\begin{figure}
\begin{center}
\epsscale{1.2}
\plotone{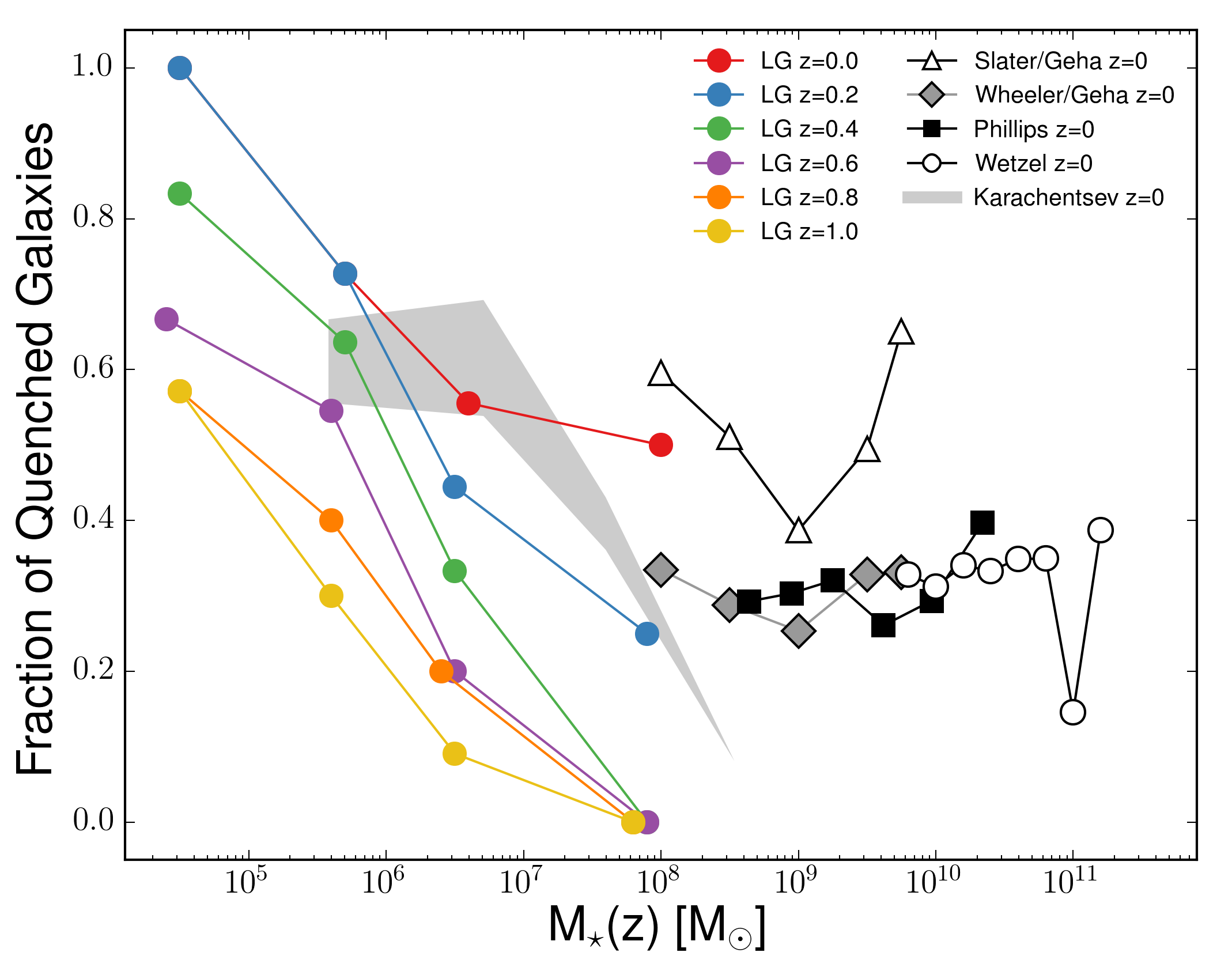}
\caption{The evolution of the quenched fraction of satellite galaxies over $\sim$ 7 dex in stellar mass.  The LG data is the same as in Figure \ref{fig:quench_zred}, while the literature values represent non-LG galaxies by \citet{geha2012, karachentsev2013, wetzel2013, slater2014, wheeler2014, phillips2015}. The grey envelope for the \citet{karachentsev2013} data shows the range of quenched fractions for whether dTrans are considered quenched or not. For M$_{\star}$ $\lesssim$ 10$^8$ \msun, there is a clear increase in quenched fraction toward lower masses, whereas above this mass the quenched fraction is roughly constant, although the analysis of \citet{slater2014} suggesting an upturn is possible.  This suggests that satellites with M$_{\star}$ $\gtrsim$ 10$^8$ \msun\ are not resistant to quenching, while those of decreasing mass are more vulnerable to quenching mechanisms that operate in high density environments (e.g., ram pressure, tidal stripping) and possibly reionization at the lowest masses.}
\label{fig:quench_satellite}
\end{center}
\end{figure}

Aside from physical effects, we must also consider the possibility of systematic uncertainties in the \citet{karachentsev2013} data.  In particular, inaccurate morphological classifications at lower masses, incompleteness, and/or inaccurate distances could all influence the quenched fraction beyond the nominal Poisson errors.  We address this confusion by computing the quenched fractions in the cases that all dTrans and quenched/not quenched, and plot the range as envelopes in Figure \ref{fig:quench_zred}.  While these results are intriguing, it is clear that a more detailed study of low-mass field galaxies in the very nearby universe will help clarify the origin of upturn in the \citet{karachentsev2013} data.

The redshift evolution in Figure \ref{fig:quench_zred} is consistent in a picture of essentially monotonic quenching.  That is, once a galaxy becomes quenched, it is unlikely to go back to being star-forming.  There are many mechanisms that can induce and maintain a galaxy's quenched status (e.g., feedback, ram pressure), but essentially only gas accretion can re-ignite star formation.  Given that this single process must compete with a plethora of quenching mechanisms, it is unlikely to observe significant decreases in the quenched fraction toward the present.  Indeed, this trend is observed in both the LG and \citet{wetzel2013} data.

Applying this same reasoning to the $z=0$ literature data, we can conclude that galaxies with M$_{\star}$$\sim$10$^8$-10$^{10}$ \msun\ have had similarly low quenched fractions in the past.  That is, galaxies in this mass range truly appear to be resistant to quenching, irrespective of their environment.

\subsection{Satellite Quenching}
\label{sec:satellitequench}

Although we have just discussed the LG in the context of all galaxy quenching (i.e., centrals and satellites), a more conventional approach is to analyze only centrals or only satellites.  Thus, in Figure \ref{fig:quench_satellite}, we plot the redshift evolution of our LG data along with satellite quenched fractions from several $z=0$ satellite quenching studies. 

The overall finding of these studies is that satellites with M$_{\star} \gtrsim$ 10$^{8}$ \msun\ have nearly constant quenched fractions over over a large range of stellar mass (M$_{\star} \sim$ 10$^{8}$ - 10$^{11}$ \msun).  There are some minor inconsistencies in trends reported by \citet{slater2014} and \citet{wheeler2014}, which applied different selection criteria to the same \citet{geha2012} dataset. This disagreement highlights another challenge in quantitatively interpreting data on quenching from different studies, a point that can be extended to our derivation of the satellite quenched fraction from the \citet{karachentsev2013} catalog.

Despite subtle catalog differences, the general belief is that satellites with M$_{\star} \gtrsim$ 10$^{8}$ \msun\ are fairly resistant to quenching, even when acted on by ram pressure and/or tidal effects.  The exact timescales of quenching are still sensitive to choices in running and interpreting simulations.  However, commonly referenced timescales for becoming quenched after a satellites enters the virial radius of a massive galaxies are $>$5-9 Gyr \citep{slater2014, wheeler2014, phillips2015}.  Although \citet{slater2014} suggest that following a pericentric passage, quenching must have more rapid timescales of $\sim$ 1-2 Gyr, which is consistent with our current understanding of Leo {\sc I}.  We discuss more details on the relationship model-based infall times and SFH-based quenching epochs in the next section.

For galaxies with M$_{\star} \lesssim$ 10$^{8}$ \msun, the increase in quenched fraction with decreasing mass suggests that lower mass systems are more vulnerable to the environment quenching effects in a group environment.  However, a comparison with the \citet{karachentsev2013} field data points in Figure \ref{fig:quench_zred} suggests that secular processes (e.g., feedback) may also be important.  It is unclear how to reconcile these two figures in light of the current paradigms about gas-loss in low-mass galaxies, which do not make a clear prediction for mass-dependent quenching in the field.  While this observation may hint at new insight into the physics of quenching in low-mass galaxies, it also is clouded by potential selection effects and small number statistics as discussed in \S \ref{sec:ushape}.  It is further complicated by our primitive understand of the effects of reionization on low-mass galaxies \citep[e.g,.][]{benitezllambay2014, onorbe2015}.  While many in this sample are above the stellar mass range assumed to be affected by reionization, a new generation of low-mass galaxy simulations suggests that the effects of reionization may extended to higher mass systems \citep[e.g.,][]{benitezllambay2014}.  In such cases, it may be that reionization in tandem with other effects (e.g., feedback) can quench or strongly suppress star formation in galaxies more massive than posited by the first generation of models \citep[e.g.,][]{ricotti2005}. For deeper empirical and theoretical understanding, more extensive surveys of low-mass galaxies in isolated environments are needed to understand the significance of these trends revealed in this paper.

\section{Quenching vs. Infall Timescales}
\label{sec:lgmodels}

Constraints on the processes responsible for quenching and its efficiency can come from comparing the infall time and quenching time.  The `delay time' between the two is a proxy for how long a galaxy can retain its gas and continue forming stars after entering the virial radius of a massive host.  Alternatively, if quenching happens prior to infall, this may indicate that non-environmentally driven processes (e.g., reionization) were responsible.  

\begin{figure}
\begin{center}
\epsscale{1.2}
\plotone{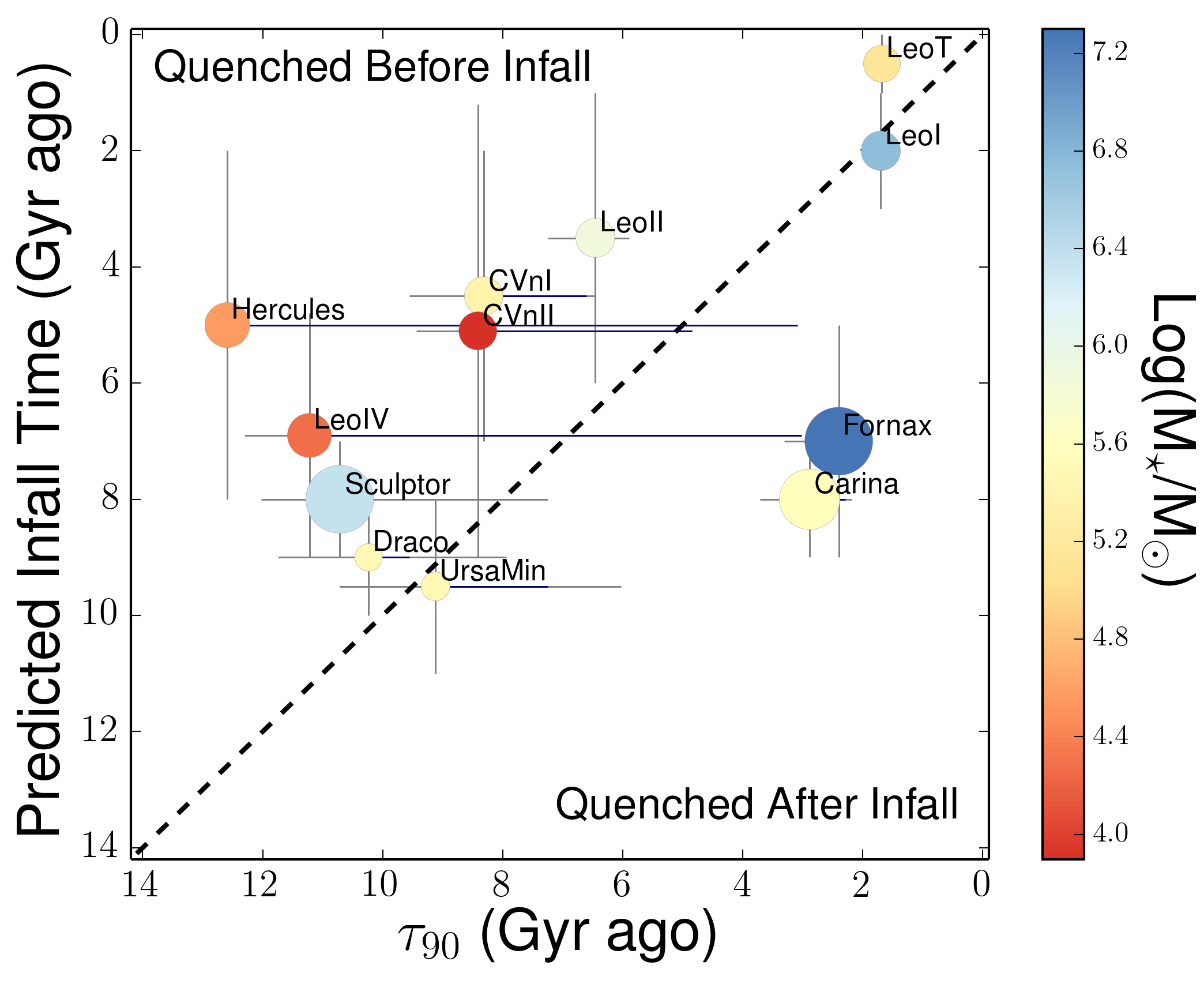}
\caption{Constraints on the delay time (the difference between infall time and the quenching epoch) for select galaxies associated with the MW.  The infall times on the $y$-axies are based on analysis of Via Lactea presented in \citet{rocha2012} and our values of \tauninety\ are on the $x$-axis.  Point sizes are proportional to the half-light radii, and error bars represent the total (grey) and random uncertainties (navy; in the case of \tauninety).  Galaxies located to the upper left quenched prior to infall, while those in the lower right quenched after infall.  Galaxies that quenched after infall appear to have done so within $\sim$ 1-4 Gyr after entering the virial radius of the MW. }
\label{fig:rochas_compare}
\end{center}
\end{figure}

While quenching epochs come directly from our SFH measurements, infall times cannot be measured directly and instead must be inferred from models.  A common approach is to match sub-halos from N-body simulations with observed galaxies by comparing their masses (via abundance matching) and available phase space information (e.g., tangential and radial velocities).  \citet{rocha2012} provide a clear example of this type of modeling applied to galaxies associated with the MW.  Using Via Lactea {\sc II} \citep{diemand2008}, they quantify the correlation between the present day orbital energy and the epoch at which a galaxy most recently entered the virial radius of the MW.  Combined with available radial and proper motions (when available), they predict the infall time probability distribution functions for a dozen dwarf galaxies associated with the MW.  In the remainder of this section, we compare their infall time predictions with our quenching timescales and explore implications for quenching mechanisms.

Figure \ref{fig:rochas_compare} shows several interesting trends in the comparison between \tauninety\ and the predicated infall times.  First, several of the lowest mass galaxies appear to have quenched before infall.  For example, Hercules quenched $\sim$ 12 Gyr ago, but only entered the virial radius of the MW $\sim$ 5 Gyr ago, suggesting that it spent the majority of its life away from the environmental influence of the MW.   Six other galaxies (Leo {\sc IV}, CVn {\sc I}, CVn {\sc II}, Leo {\sc II}, Sculptor, and Draco) also have earlier quenching  than infall times.  Of these six galaxies, more than half (Hercules, Leo {\sc IV}, CVn {\sc I}, and CVn {\sc II}) have been previously identified as putative `fossils of reionization', i.e., their star formation was quenched due by reionization \citep[e.g.,][]{bovill2011b, brown2012, brown2014}.  If they were located outside the virial radius of the MW prior to quenching, the likelihood of quenching due to environmental processes (e.g., ram pressure) seems drastically lower, increasing the likelihood that reionization played a significant role in quenching their star formation.  

Second, at the other extreme, Carina and Fornax appear to have quenched several Gyr after their infall times. Figure \ref{fig:rochas_compare} suggests a delay of $\sim$ 5 Gyr between when these galaxies fell in, and when their star formation stopped.  This scenario would be qualitatively similar to the `delayed-then-rapid' quenching scenario for more massive satellites (M$_{\star}$ $\gtrsim$ 10$^9$ - 10$^{10}$ \msun) that was proposed by \citet{wetzel2013}.  In this scenario a satellite continues to form stars for $\sim$ 2-4 Gyr after infall, before rapidly quenching in less than 1 Gyr. We will explore more detailed modeling of this scenario in a future paper.

Third, there are a handful of galaxies that quenched near or shortly after their infall time.  One example is Leo {\sc I}. Leo {\sc I} appears to have fallen into the MW's halo $\sim$2.3 Gyr ago \citep{rocha2012, sohn2013}.  $\sim$ 1 Gyr ago, Leo {\sc I} appears to have reached pericenter at $\sim$ 90 kpc from the MW, which is coincident with its value of \tauninety.  This suggests an extremely small delay time, which is in stark contrast to the comparably luminous MW dSph Fornax.  However, Leo {\sc I} is believed to be on a highly inclined orbit, while Fornax is more circular, suggesting that exact orbital trajectory may play a role in a the delay time between infall and quenching. 

Carina, which has a lower stellar mass than Fornax, also quenched $\sim$ 5 Gyr after infall, supporting the idea that mass may not be the sole determinant in quenching timescale, and exact orbital history may matter.  Interestingly, in the case of Carina, there is an open question as to whether its SFH shows multiple distinct bursts followed by periods of little or no star formation \citep[e.g.,][]{smeckerhane1994, hurleykeller1998, bono2010}.  Interpreted in the context of infall, it could be that Carina has made more than one passage around the MW, but only lost all of its gas and became completely quenched within the last few Gyr.  As shown in \citet{weisz2014a}, our SFH of Carina has similar bursts and lulls in its SFH to previous literature derivations.  However, when all uncertainties are considered, it is possible that the SFH was never completely quiescent between bursts.

Finally, \citet{rocha2012} suggest that Leo~T only entered MW virial radius within the last Gyr.  Observationally, the impact of this recent accretion appears to be minimal.  Leo~T has a large baryonic gas fraction (0.8) and displays spatially symmetric \hi\ and stellar distributions \citep[e.g.,][]{irwin2007, ryanweber2008}. Further, Leo~T has nearly constant star formation over the last few Gyr, showing little evidence for Leo {\sc I}-like enhancement upon infall.  It does lack detectable \halpha\ emission, which could be interpreted as a break in the SFH.  However, it could also be due to stochastic sampling of the stellar IMF and/or cluster mass function, which is believed to be common in such low-mass systems \citep[e.g.,][]{fumagalli2011, dasilva2012}.

\section{Summary}
\label{sec:summary}

We have used the SFHs of 38 LG dwarf galaxies to explore the quenching characteristics of low-mass galaxies (10$^4$ $\lesssim$ M$_{\star}$ $\lesssim$ 10$^8$ \msun).  The SFHs were uniformly measured from the analysis of resolved star CMDs based on HST/WFPC2 imaging that were presented in \citet{weisz2014a}.  Using a proxy for the quenching epoch as the lookback time which 90\% of a galaxy's stellar mass formed, we found the following results:

\begin{enumerate}
\item Lower mass galaxies tend to quench prior to higher mass galaxies.  Specifically, the quenched fraction of the lowest mass galaxies (\logavgmass$=$4.5) is 0.4 at $\sim$ 10 Gyr ago and 1.0 at $\sim$ 3-4 Gyr ago.  In contrast, at the same epochs, the highest mass systems (\logavgmass$=$ 8.0) have a quenched fraction of $<$ 10\% and $\sim$ 40\%.  The quenched fraction appears to vary smoothly as a function of mass and lookback time between these extremes.  Due to incompleteness of our sample, the quenched fraction of the lowest mass galaxies may be larger at early times.  However, several of the lower mass dSphs in the LG show evidence for extended SFHs, so the assumption that all low-mass galaxies quenched early may not be entirely accurate.

\item Within \rvir\ of either the MW or M31, there is little correlation between present distance to a host and quenching epoch. Outside of \rvir\ the quenched fraction is effectively zero at all lookback times.

\item There are hints of systematic differences in the quenching epochs of M31 and MW dSphs.  The quenched fraction of M31 dSphs jumps from 0.2 at $\sim$ 7 Gyr ago to 1.0 by $\sim$ 5 Gyr ago, whereas the quenched fraction of MW dSphs steadily increases over cosmic time.  Further, the four M31 dEs all appear to have quenched between $\sim$ 3 and 4 Gyr ago.  However, the sample size and shallow CMDs of the M31 companions make it challenging to draw any definitive conclusions.

\item  There are clear trends in the quenched fraction over $\sim$ 7 dex in stellar mass.  The lowest (M$_{\star}$ $\lesssim$ 10$^{5}$ \msun)  and highest mass (M$_{\star}$ $\sim$ 10$^{11.5}$ \msun) galaxies always have the highest quenched fractions between $z=0-1$, while galaxies with M$_{\star}$ $\sim$ 10$^8$ - 10$^{10}$ \msun\ always have the lowest quenched fraction. Given that the quenched fraction essentially monotonically increases toward the present, we conjecture that galaxies with M$_{\star}$ $\sim$ 10$^8$ - 10$^{10}$ \msun\ are universally the most difficult to quench.  

\item A comparison with the predicted infall times from \citet{rocha2012} for select galaxies associated with the MW allows us to estimate delay times (i.e., the time between infall and quenching).  We find that lower mass systems tended to have quenched prior to infall, while higher mass systems quenched after infall, and exhibit delay times of $\sim$ 1-4 Gyr.

\end{enumerate}

\section*{Acknowledgements}

The authors would like to thank the anonymous referee for insightful comments that helped increase the scope and improve the clarity of this paper.  DRW would like to thank Mike Boylan-Kolchin, Charlie Conroy, Eric Bell, Colin Slater, Andrew Wetzel, and Erik Tollerud for insightful discussion about quenching and low-mass galaxies; Marla Geha, Coral Wheeler, John Phillips, and Andrew Wetzel for making their data available for comparison; and Hans-Walter Rix and the MPIA for their hospitality during the assembly of this paper.  Support for DRW is provided by NASA through Hubble Fellowship grants HST-HF-51331.01. Additional support for this work was provided by NASA through grant number HST AR-9521 from the Space Telescope Science Institute, which is operated by AURA, Inc., under NASA contract NAS5-26555. EDS and DRW were supported in part by the National Science Foundation under Grant No. PHYS-1066293 and the hospitality of the Aspen Center for Physics. This research made extensive use of NASA's Astrophysics Data System Bibliographic Services.  In large part, analysis and plots presented in this paper utilized IPython and packages from NumPy, SciPy, and Matplotlib \citep[][]{hunter2007, oliphant2007, perez2007, astropy2013}.

\bibliographystyle{apj}
\bibliography{lg_quenching.revise2.bbl}

\end{document}